\documentstyle[12pt]{article}
\newcommand{\half}{{1\over2}}
\newcommand{\be}{\begin{equation}}
\newcommand{\ee}{\end{equation}}
\newcommand{\bea}{\begin{eqnarray}}
\newcommand{\eea}{\end{eqnarray}}
\newcommand{\none}{\nonumber \\}
\newcommand{\req}[1]{Eq.(\ref{#1})}

\newcommand{\g}{\overline{g}}
\newcommand{\gbar}{\overline{g}}
\newcommand{\phibar}{\overline{\phi}}
\newcommand{\V}{\overline{V}}

\newcommand{\jbar}{\overline{j}}
\newcommand{\k}{\overline{K}}

\newcommand{\x}{\prime }
\newcommand{\y}{\lambda }

\begin{document}
\begin{center}
{\bf Quantum Corrections to the Thermodynamics of\\
Charged 2-D Black Holes}  
\end{center}
\vspace{0.5 cm}
\begin{center}
{\sl by\\
}\vspace*{0.50cm}
{\bf A.J.M. Medved${}^\flat$   and G. Kunstatter$^\sharp$\\}
\vspace*{0.50cm}
{\sl
$\flat$ Dept. of Physics and Winnipeg Institute for 
Theoretical Physics, University of Manitoba\\
Winnipeg, Manitoba\\
Canada R3T 2N2\\
{[e-mail: joey@theory.uwinnipeg.ca]}}\\
\vspace*{0.50cm}
{\sl
$\sharp$ Dept. of Physics and Winnipeg Institute for Theoretical
Physics\\
 University of Winnipeg, Winnipeg, Manitoba\\
Canada R3B 2E9\\
{[e-mail: gabor@theory.uwinnipeg.ca]}
}
\end{center}
\bigskip\noindent
{\large ABSTRACT}
\par
\noindent

We consider one-loop quantum corrections to the thermodynamics of
 a black hole in generic 2-dimensional dilaton gravity. The classical
 action is the most general diffeomorphism invariant action in 1+1
 space-time dimensions that contains a metric, dilaton, and Abelian
 gauge field, and having at most second derivatives of the fields. 
 Quantum corrections are introduced by considering the effect of 
 matter fields conformally coupled to the  metric and 
non-minimally coupled to the dilaton.
 Back reaction of the matter fields ( via non-vanishing trace conformal
 anomaly ) leads to quantum corrections to the black hole geometry.
 Quantum corrections also lead to modifications in the gravitational
 action and hence in expressions for thermodynamic quantities. One-loop
 corrections to both geometry and thermodynamics ( energy, entropy )
 are calculated for the generic theory. The formalism is then
 applied to a charged black hole in spherically symmetric gravity and to
 a rotating BTZ black hole.

\newpage

\section{Introduction}\medskip
\par

There exists a strong analogy between the properties of black holes and
 conventional thermodynamical systems \cite{thermo}. In this analogy the entropy
 of the black hole is directly proportional to the surface area of its
 event horizon and literature refers to this quantity as the Bekenstein-Hawking
 entropy \cite{bekenstein}. Meanwhile, the temperature of the black hole is proportional
 to the surface gravity of its horizon. In spite of this established
 correspondence, there is still a lack of understanding of precisely what
 accounts for black hole entropy which is a pure geometrical quantity.
 In a usual statistical mechanical system the entropy is explained in
 terms of the degrees of freedom of its microscopic constituents. However
 a black hole has a limited number of such degrees of freedom,
 as demonstrated by
 the so called "No-Hair" theorems \cite{hair}.

In recent literature, varied attempts have been made to derive black hole
 entropy on statistical mechanical principles, with varying
 degrees of success \cite{stringy}\cite
{carlip}\cite{induced}\cite{loop}. For instance, Strominger and Vafa
 \cite{stringy} counted the degeneracy of soliton bound states for extremal black
 holes in string theory. In a very different and more geometrical approach,
 Carlip \cite{carlip} counted horizon edge states in a gauge theory formulation
 of 2+1 anti-deSitter gravity. Another approach that has been investigated
 involves Sakharov's theory of induced gravity \cite{sak} following a proposal
 by Jacobson \cite{jac}. Recent work along these lines to generate black hole
 entropy has been done by Frolov, Fursaev, and Zelnikov \cite{induced}. 

 The success of a number of very diverse approaches seems to suggest that
the correct  explanation for  black hole entropy may in some sense
 be universal. That is, it should 
 depend explicitly on neither the macroscopic gravitational form nor
 on a hidden microscopic quantum theory \cite{wald}. Consequently, it may prove
 beneficial to study as wide a range of theories as feasible and in doing
 so look for model independent features. Such observations could
 potentially provide valuable insight as to the geometrical origins
 of black hole entropy. To this end, we examine the thermodynamic
 properties of black holes in generic dilaton gravity coupled to an
 Abelian gauge field in 1+1 dimensions. This provides a very extensive
 class of models which allow for black hole solutions. Even with the
 2-dimensional limitation, many such models are seen to have direct
 physical significance. For instance, in the spherically symmetric
 reduction of 4-dimensional Einstein-Maxwell gravity the 
 dilaton scalar field corresponds to 
 radial distance.  Also it has been shown that black hole solutions
 of constant curvature gravity in 2-dimensions ( Jackiw-Teitelboim\cite{JT}) are
 in fact projections of BTZ black holes described by 2+1 gravity with
 axial symmetry \cite{ortiz}.

In recent work we have studied the  classical 
thermodynamic properties of generic
 dilaton gravity via a Hamiltonian partiton function method
 \cite{star}\cite{shelemy}.  In the present paper we calculate
 the thermodynamics so as to include one-loop corrections. The approach
 we use here is based on York's Euclidean-action method  
\cite{york}\cite{brown} which
 in turn follows from the Gibbons-Hawking path integral
 formalism \cite{gibbon}. This
 entails taking the black hole to be in a state of thermal equilibrium
 with evaporated radiation and then relating the periodicity of
 Euclidean time with the inverse thermodynamic temperature. First-order
 quantum corrections are introduced into the procedure  via a technique
 applied by Frolov, Israel, and Soldukin \cite{frolov}  in the study of spherically
 symmetric charged black holes. The basic idea is to add to the classical 
action a correction corresponding to the one-loop effective action obtained by
integrating out matter fields coupled to the metric and non-minimally coupled to the dilaton.
This one loop effective action is a suitable generalization\cite{odintsov1,hawk,kummer1,dowk} of the 
Polyakov action obtained from the 2-D conformal (or trace) anomaly\cite{trace}. The effect of these quantum corrections on the black hole thermodynamics is two-fold. First, it modifies the black hole geometry due to the
non-vanishing one-loop  effective stress-energy tensor. Secondly, the  surface terms which give rise to the black hole free energy also acquire quantum corrections. As a result the formulae relevant to calculating thermodynamic
 quantities ( energy and entropy ) are modified as well.  Our results will
  hopefully provide insight into the nature of such corrections for generic
 dilaton gravity as well as provide the template for closer examination
 of a myriad of specific theories.

This paper is arranged as follows. In Section 2 we introduce the action for
 generic 2-dimensional dilaton gravity coupled to an Abelian gauge field.
 Here we present the most general solution to the field equations and
 by way of the Euclidean action approach \cite{york}  we are able to describe
  black hole thermodynamic properties at a classical level. In Section 3
 we introduce one-loop quantum corrections due to matter fields propogating
 on a curved background. The resulting modifications to the  black hole
 geometry are deduced by applying the formalism of Frolov et al. \cite{frolov}.
 In Section 4 we calculate the quantum corrections to black hole energy
 and entropy. In Sections 5 and 6 we apply our results to the specific
 examples of  charged black holes in spherically symmetric gravity and 
 rotating BTZ black holes respectively. \footnote{Quantum corrections to the thermodynamics of the BTZ black hole and some classes of 2d charged black holes have been previously considered using different methods in \cite{odintsov2}
while  \cite{bytsenko} has examined quantum gravitational corrections to
the entropy of the BTZ black hole.} For simplicity cases consider
 minimal coupling of matter fields with the dilaton however the formalism
 to be presented is readily extendable to more general coupling scenarios.
 Section 7 summarizes the paper and considers future prospects for
 related work.

\section{Classical Theory}

In two spacetime dimensions the Einstein tensor vanishes
identically.  Consequently, the
construction of a dynamical theory of gravity with no more than two
derivatives of the metric
in the action requires the introduction of a scalar field, namely
the dilaton.  Recent works
have demonstrated that the dilaton is more than a lagrange
multiplier but significant in
determining both the symmetries and topologies of the solution
 \cite{DGK}.

Here we consider the most general Lorentzian action functional
depending on the metric
tensor $g_{\mu\nu}$, dilaton scalar field $\phi $ and Abelian gauge
field $A_{\mu\nu}$ in two
spacetime dimensions \cite{banks}\cite{DK}:
\bea
W[g,\phi ,A] &=& {1\over{2G}}\int
d^2x\sqrt{-g}\left[D(\phi )R(g)\right.\nonumber\\ 
&&\quad +
\left.{1\over2}g^{\alpha\beta}\partial_{\alpha}\phi\partial_{\beta
}
\phi + {1\over{l^2}}V(\phi ) - {2G\over 4}Y(\phi
)F^{\alpha\beta}
F_{\alpha\beta}\right]
\label{eq: 1}
\eea
where $G$ is the dimensionless 2-d Newton constant, $F_{\mu\nu} = 
\partial_{\mu}A_{\nu} - \partial_{\nu }A_{\mu }$ ,and l is a fundamental
constant of dimension length. 
Also
$D(\phi ), V(\phi ),$ and $Y(\phi )$ are arbitrary functions of the
dilaton field.  

Variation of the action with respect to the metric, dilaton field
and gauge field respectively
leads to the following set of field equations.

\bea
-2\nabla_{\alpha }\nabla_{\beta }D(\phi ) & + & \nabla_{\alpha
}\phi\nabla_{\beta }
\phi + g_{\alpha\beta}\left(2\Box D(\phi )-\half(\nabla\phi
)^2-{1\over{l^2}}
V(\phi )\right.\nonumber\\
 & + & \left.{G\over 2}Y(\phi )F^{\mu\nu}F_{\mu\nu}\right)-2G
Y(\phi )F^{\gamma}_{\alpha }F_{\beta\gamma} = 0
\label{eq: 2}
\eea
\be
-\Box\phi + 
\left[R{\delta D\over\delta\phi }+{1\over{l^2}}
{\delta V\over{\delta\phi}}-{G\over 2}{\delta Y\over \delta\phi }F^
{\alpha\beta }F_{\alpha\beta }\right] = 0
\label{eq: 3}
\ee
\be
\nabla_{\beta }(Y(\phi )F^{\alpha\beta}) = 0
\label{eq: 4}
\ee
Directly solving Maxwell's equation \req{eq: 4} yields
\be
F = {{\sqrt{-g}q}\over {Y(\phi )}}
\label{eq: 5}
\ee
where $F$ is defined implicitly by $F_{\mu r} = F\epsilon_{\mu\nu}$
and
$q$ is a constant that
corresponds to Abelian charge.  Next we define an ``effective''
potential 
${\tilde V}(\phi ,q)$ such that
\be
{\tilde V}(\phi ,q) = V(\phi ) - Gl^2{q^2\over {Y(\phi )}}
\label{eq: 6}
\ee
The action and remaining field equations Eqs.[\ref{eq: 2},
\ref{eq: 3}] can be rewritten as
follows:
\be
W[g,\phi ,q] = {1\over {2G}}\int d^2x\sqrt{-g}\left[D(\phi
)R(g) + \half 
(\nabla\phi )^2 + {1\over {l^2}}\tilde{V}(\phi ,q)\right]
\label{eq: 7}
\ee
\be
-2\nabla_{\alpha }\nabla_{\beta}D(\phi ) +
\nabla_{\alpha}\phi\nabla_{\beta}
\phi + g_{\alpha\beta}\left(2\Box D(\phi ) - \half 
(\nabla\phi )^2 - {1\over{l^2}}{\tilde V}(\phi , q)\right) =
0
\label{eq: 8}
\ee

\be
-\Box\phi + 
R{{\delta D}\over{\delta\phi }} + {1\over l^2}
{\delta{\tilde V}\over {\delta\phi }} = 0
\label{eq: 9}
\ee

Obtaining the solutions for an action of this form has been well
documented in prior works  \cite{star}\cite{DK} so only a brief account will be
presented here.  First the action is reparameterized   
thereby eliminating the kinetic term (requires $D(\phi )
\neq 0$ and ${{dD}\over {d\phi }}\neq 0$ for any admissable value
of $\phi $ ).
\be
\phibar = D(\phi )
\label{eq: 10}
\ee
\be 
\Omega^2 = \exp\half\int{{d\phi }\over {dD/d\phi }}
\label{eq: 12}
\ee
\be
\g_{\mu\nu} = \Omega^2(\phi )g_{\mu\nu}
\label{eq: 11}
\ee
\be
{\overline{V}}(\phibar ,q) = {\tilde V}(\phi ,q)/\Omega^2(\phi )
\label{eq: 13}
\ee
The reparameterized action is then as follows:
\be
W[\gbar ,\phibar ,q] = {1\over {2G}}\int d^2x\sqrt{-\gbar}
\left[\phibar R(\gbar ) + {1\over {l^2}}\V
(\phibar,q)\right]
\label{eq: 13.5}
\ee

  The timelike killing
vector for the resultant field equations is easily identifiable. 
It is found to be:
\be
\k^{\mu } = {l\epsilon^{\mu\nu}\over \sqrt{-\gbar}}\partial_{\nu
}\phibar
\label{eq: 14}
\ee
With norm given by
\be
|\k|^2 = \jbar(\phibar ) - 2GlM
\label{eq: 15}
\ee
where
\be
\jbar(\phibar ) = \int^{\phibar }_od\phibar\V(\phibar ,q)
\label{eq: 16}
\ee
and $M$ is a constant of integration identified as the mass
observable.

Next we choose a local coordinate system in which $\phibar $ and hence
$\phi $ have spatial dependence only. The final solutions in these
static coordinates are then
obtained by exploiting the form of the killing vector.  These are
found to be:
\be
{\overline {\phi }} = {x\over l}
\label{eq: 16.25}
\ee
\be
ds^2 = -\gbar(x)dt^2 + \gbar^{-1}(x)dx^2
\label{eq: 16.5}
\ee
where
\be
{\overline{g}}(x) = \jbar(\phibar ) - 2GlM
\label{eq: 16.75}
\ee
We can then re-express this solution in terms of the original
parameterization as follows:
\be
\phi = D^{-1}({x\over l})
\label{eq: 17}
\ee
\be
ds^2 = -g(x)dt^2 + g^{-1}(x)\Omega^{-4}(\phi (x))dx^2
\label{eq: 18}
\ee
where
\be
g(x) = {1\over \Omega^2(\phi (x))}\left[\jbar({x\over
l})-2GlM\right]
\label{eq: 19}
\ee

The necessary condition for a given theory to admit black hole
configurations is the existence of apparent horizons. That is,
 spacetime curves of the form
$\phi (x,t) = \phi_o$ (constant) where $\phi_o$ satisfies $g(\phi_o
; M, q) = 0$.  The nature of a given black hole solution can
be revealed by considering $dg / d\phi $ evaluated at these
event horizons $\phi = \phi_o$.  For a fixed value of mass M there
may exist critical values
of charge $q (M)$ so that
this derivative vanishes.  For such critical values the function 
$g(\phi ; M, q)$ may have either a local extremum or point of
inflection at the horizon.  If it is an extremum the norm of the
killing vector does not change sign when passing through the event
horizon.   As $q$ is varied away from its critical value either the
horizon will disappear or two event horizons (inner and outer) will
appear.  The latter case signifies the presence of an extremal
black hole when $q$ is at its critical value.  For a point of
inflection the norm of the killing vector does change sign but as
$q$ is varied away from its critical value one expects the
formation of either one or three horizons \cite{DK}.

For the  subsequent (thermodynamic) analysis we consider the Euclidean
sector such that $t\rightarrow it$.  Hence we re-write the action
\req{eq: 7}
with respect to the Euclidean metric tensor:
\bea
 W_{E} &=& -{1\over 2G}\int d^2x\sqrt{g}\left[D(\phi )R(g) +
{1\over 2}(\nabla\phi )^2\right.\nonumber\\
&&\quad + \left.{1\over l^2}{\tilde V}(\phi ,q)\right]-{1\over G}
\oint_{outer\ boundary} dt\gamma D(\phi)\nabla_{\alpha }
n^{\alpha }
\label{eq: 19.5}
\eea
(Henceforth the subscript E on the Euclidean action will be implied).
The second integral in \req{eq: 19.5} is the extrinsic curvature
boundary term. It is included so that when second derivatives of the metric are
cancelled off (via appropriate integration by parts) then the resulting
 total divergences on the outer boundary will be cancelled off as well
 \cite{gibbon}. Here we
define $n_{\mu }$ as the outward unit vector normal to the outer
boundary enclosing the black hole and $\gamma$ as the induced
metric appropriate for evaluating the line integral.

We re-write the Euclidean static metric in the following form
\be
ds^2 = g(x)dt^2 + e^{-2\lambda (x)}g^{-1}(x)dx^2
\label{eq: 20}
\ee
where $e^{\lambda(x)} = \Omega^2(\phi (x))$.  For this metric it
follows that
\be
\sqrt{g}= e^{-\lambda (x)}
\label{eq: 21}
\ee
\be
R = -e^{\lambda (x)}\left(e^{\lambda
(x)}g^{\prime}(x)\right)^{\prime}
\label{eq: 22}
\ee
where $\prime $ indicates differentiation with respect to $x$.  The
coordinates $t, x$ describe a disc and will be taken to range
between the limits $x_+\leq x \leq L$ and $0 \leq t \leq 2\pi\beta
$.  Here $x = x_+$ represents the black hole horizon (ie. - $g(x_+)
= 0), x = L$ is the outer boundary of the black hole (ie. - box
size), and $\beta$ is the asymptotic inverse temperature.  It can
be
shown that regularity of the solution requires the absence of a conical
singularity which leads to the following condition: 
\be
\beta = {2e^{-\lambda (x_+)}\over g^{\prime }(x_+)}
\label{eq: 23}
\ee
Note that application of standard thermodynamics requires using the 
inverse temperature of the box
${\overline {\beta }}$  which is "red-shifted" from the previously
defined quantity such that:
\be
{\overline{\beta }} = g^{1/2}(L)\beta
\label{eq: 24}
\ee
For this metric it also follows that the extrinsic curvature
(defined by the boundary term of the action) can
be expressed \cite{frolov}:
\be
\gamma\nabla_{\alpha }n^{\alpha} = \half e^{\lambda (L)}g^{\prime
}(L)
\label{eq: 25}
\ee

Using these results Eqs.[\ref{eq: 20}-\ref{eq: 25}] we can
express the Euclidean action functional (\req{eq: 19.5}), with respect to the
generic static metric giving:
\bea
W &=& {-\pi\beta\over G}\int^L_{x_+}dx\left(D^{\prime }
(x)e^{\lambda (x)}g^{\prime }(x) + {e^{\lambda (x)}\over 2}g(x)
(\phi^{\prime }(x))^2\right.\nonumber\\
&&\quad + \left.{e^{-\lambda (x)}{\tilde V}(x)\over l^2}\right)
-{2\pi\over G}D(x_+)
\label{eq: 26}
\eea
  We can use this form of the action to
derive thermodynamic
properties of interest.  These include the free energy $F =
(2\pi{\overline{\beta }})
^{-1} W $, energy $E = (2\pi )^{-1}\partial_{\overline
{\beta }} W $,
and entropy $S = ({\overline {\beta }}\partial_{\overline {\beta
}}-1) W $.
\bea
F &=& -{1\over 2Gg^{1/2}(L)}\int^L_{x_+}dx\left(D^{\prime }(x)e^
{\lambda (x)}g^{\prime }(x) + {e^{\lambda (x)}\over 2}g(x)
(\phi^{\prime }(x))^2\right.\nonumber\\
&&\quad + \left.{e^{-\lambda (x)}{\tilde V}(x)\over l^2}\right) 
- {D(x_+)\over G{\overline{\beta}}}
\label{eq: 27}
\eea
\be
E = -{1\over 2Gg^{1/2}(L)}\int^L_{x_+}dx\left(D^{\prime }(x)e^{\lambda
(x)}g^{\prime }(x) + {e^{\lambda (x)}\over 2}g(x)(\phi^{\prime }
(x))^2 + {e^{-\lambda (x)}\over l^2}{\tilde V}(x)\right)
\label{eq: 28}
\ee

\be
S = {2\pi\over G}D(x_+)
\label{eq: 29}
\ee
Since the box temperature is taken to be $T = 2\pi
{\overline{\beta }}$ from Eqs.[\ref{eq: 27}-\ref{eq: 29}]
we obtain the result $F = E-TS$. At the extremum of free energy (or
equivalently action) $\delta F = 0$ and hence the second law of
thermodynamics immediately follows.

It is possible and convenient to re-express the action (\req{eq:
26}) in a form which, except for surface terms, vanishes on shell. 
Defining  $G_{\alpha\beta }$ to be the left hand side of
\req{eq: 8} and re-writing with respect to the
coordinate system defined by metric \req{eq: 20} yields:
\bea
G_{\alpha\beta } &=& -2\delta^x_{\alpha }\delta^x_{\beta }
D^{\prime\prime}(x) + 2\Gamma^x_{\alpha\beta }D^{\prime }(x)
 + g_{\alpha\beta }\left[2e^{\lambda (x)}\left(e^{\lambda (x)}g(x)
D^{\prime }(x)\right)^{\prime }\right.\nonumber\\
&&\quad - \left.\half g(x)e^{2\lambda (x)}(\phi^{\prime }
(x))^2-{1\over l^2}{\tilde V}(x)\right]=0
\label{eq: 30}
\eea
In the case where both tensor indices represent time coordinate
denoted by 0 (and note that $\Gamma^x_{00} = -\half
e^{2\lambda}gg^{\prime })$ we get:

\bea
G^0_0 &=& -e^{\lambda(x)}\left[-e^{\lambda (x)}g^{\prime
}(x)D^{\prime }(x) - 2g(x)\left(e^{\lambda (x)}D^{\prime }(x)\right)^{\prime
}\right.\nonumber\\
&&\quad +\left. \half g(x)e^{\lambda (x)}(\phi^{\prime }(x))^2 +
{e^{-\lambda (x)}\over l^2}{\tilde V}(x)\right]
\label{eq: 31}
\eea
Using this result to substitute for the second and third terms in
the integrand of \req{eq: 26} : 
\be
 W = -{\pi\beta\over G}\int^L_{x_+}dx\left[-e^{-\lambda (x)}
G^0_0+ 2\left(g(x)e^{\lambda (x)}D^{\prime }(x)\right)^{\prime }
\right] - {2\pi\over G}D(x_+)
\label{eq: 32}
\ee
Since the second term in the integrand is a total derivative and
$g(x_+) = 0$ it follows that:
\be
 W = {\pi\beta\over G}\int^L_{x_+}e^{-\y(x)}G^0_0dx -
{2\pi\beta\over G}e^{\y(L)}g(L)D^{\x}(L) - {2\pi\over G}D(x_+)
\label{eq: 33}
\ee
Reconsider the energy $E = (2\pi )^{-1}\partial_{{\overline{\beta
}}} W $.  Since thermodynamic quantities are presumed to be
calculated for equilibrium configurations (i.e. on shell) here we
can set $G^0_0 = 0$ giving an energy which reduces to an outer
boundary surface term:
\be
E = -{1\over G}e^{\y(L)}g^{\half }(L)D^{\x}(L)
\label{eq: 34}
\ee
This expression is typically divergent as $ L\rightarrow\infty $.
(This follows from the divergence of the Euclidean action as the outer
boundary goes to infinity). To resolve this dilemma we compare the
energy of \req{eq: 34} with that of a carefully selected background
geometry \cite{hawk2}. The background metric will be taken here
to represent the asymptotic geometry of the black hole. Hence 
we define $ g_0 = lim_{L\rightarrow\infty} g(L) $ and the 
"subtracted energy" is given by:
\be
E_{sub} = {1\over G}e^{\lambda(L)}D^{\prime}(L)\left[g^{\half}_0-g^{\half}(L)\right]
\label{eq: 34.5}
\ee
We can justify this choice of background by noting the agreement between
this result with that attained for a Hamiltonian partition function
approach in a prior study \cite{star}.

\section{Quantum Corrected Black Hole Geometry}

In the path integral approach to black hole thermodynamics the
matter fields can be integrated out yielding an effective action which
depends only on fields in the classical action. Hence
one-loop quantum effects can be taken into account by adding a
quantum counterpart $\hbar\Gamma $ to the classical gravitational action
$W_{CL}$ (\req{eq: 19.5})  such that (assuming no matter coupling to
the  Abelian gauge field):
\be
W\left[g,\phi,q\right] = W_{CL}\left[g,\phi,q\right] + \hbar\Gamma
\left[g,\phi\right]
\label{eq: 35}
\ee
Variation of this complete action yields the quantum corrected
field equations which may be solved perturbatively.  Variation of
the action with respect to the metric gives us:
\be
G_{\alpha\beta }(g,\phi,q) + \hbar T_{\alpha\beta }(g,\phi) + O(\hbar^2) = 0
\label{eq: 35.5}
\ee
where $G_{\alpha\beta } = {\delta W_{CL}\over \delta g^{\alpha\beta
}}$ is again given by the left hand side of \req{eq: 8}
 and $T_{\alpha\beta } = 
{\delta\Gamma\over \delta g^{\alpha\beta }}$.

The general form of the one loop effective action  is \cite{odintsov1,kummer1}:
\be
\Gamma = {1\over 12}\int d^2x\sqrt{g}\left[aR{1\over \Box}R
+b(\phi)(\nabla\phi)^2{1\over \Box}R +c(\phi) R - \ln(\mu^2)b(\phi)
     (\nabla\phi)^2\right]
\label{eq: 36}
\ee
The first term is the usual trace anomaly that arises for minimally coupled scalars, while the next two terms are contributions to the anomaly from the
 non-minimal coupling to the dilaton. The last term contains an arbitrary scale factor, $\mu$, and  comes from the conformally invariant part of the effective action\footnote{We are grateful to S. Odintsov
for pointing out the necessity for including this term.}. It can be obtained using
a Schwinger-DeWitt type expansion\cite{odintsov2}. We will show that the 
scale factor $\mu$ does not affect the final thermodynamic quantities. $a$ is a constant (we will set $a=1$ for simplicity), while $b(\phi)$ and $c(\phi)$
are determined by the specific form of the coupling between the matter fields and the dilaton. For example, in dimensionally reduced spherically symmetric
gravity, $b$ and $c$ are constants\cite{hawk,dowk,odintsov1,kummer1}. Here we will treat them as arbitrary local
functions of the dilaton. 

 It is important for the
following thermodynamic analysis to put the non-local
expression \req{eq: 36} for $\Gamma $  in local form. We do this by introducing
a pair of scalar fields $\psi$ and $N$, and writing:
\bea
\Gamma &=& {1\over 12}\int d^2x\sqrt{g}\left[(\psi + N) R + (\nabla N)
\cdot(\nabla\psi
)+b(\nabla\phi)^2(\psi-\ln(\mu^2))+c\phi R \right] \nonumber\\ 
&&\quad +  {1\over 6}
 \oint_{outer\ boundary}dt\gamma(\psi +N+c(\phi))\nabla_{\alpha }
n^{\alpha }
\label{eq: 37}
\eea
where an extrinsic curvature surface term has been  added in analogy to
the classical case. It is straightforward to show that variation of \req{eq: 37}
yields the following field equations for the scalars:
\be
\psi = {1\over \Box } R
\label{eq: psi1} 
\ee
  and
\be 
N={1\over \Box } ( R+b(\nabla \phi )^2 )
\label{eq: N1}
 \ee
Substituting these equations back into \req{eq: 37} yields precisely \req{eq: 36}.
Note that in the 2-dimensional minimally coupled
case ($b=0$),  N reduces to $ \psi $ and only a single scalar field need
be introduced.

Before proceeding we show it is possible to solve explicitly for
$\psi (x)$ at the classical level (as is appropriate for this
analysis).  This is achieved by conformally mapping the coordinate
space described by the static Euclidean metric \req{eq: 20} to a
flat disc of radius $z_o$ 
and curvature $ R=\Box\psi $. This disc may be expressed
\be
ds^2 = e^{-\psi (z)}(z^2d\theta^2 + dz^2)
\label{eq: 38}
\ee
where the disc coordinates $\theta $ and $z$ are taken to range
between $0 \leq \theta \leq 2\pi $ and $0 \leq z \leq z_0$. 
Substituting the Euclidean static metric  \req{eq:
20} into the left hand side gives
\be
g(x)dt^2 + g^{-1}(x)e^{-2\y (x)}dx^2 = {\overline
{e}}^{\psi(z)}(z^2d\theta^2 + dz^2)
\label{eq: 39}
\ee
where $t = \beta\theta $ and $x_+ \leq x \leq L$.  The following
relations follow directly from \req{eq: 39}:
\bea
g(x)dt^2 &=&e^{-\psi }z^2d\theta^2\nonumber\\&=& 
{e^{-\psi }z^2\over \beta^2}dt^2
\label{eq: 40}
\eea
\be
g^{-1}(x)e^{-2\lambda }dx^2 = e^{-\psi
}dz^2
\label{eq: 41}
\ee
Using \req{eq: 40} to solve for $z$ and \req{eq: 41} to solve
for $dz$ gives us:
\be
z = \beta\sqrt{g}e^{+\psi /2}
\label{eq: 42}
\ee
\bea
dz&=&{e^{-\lambda}e^{+\psi/2}\over\sqrt{g}}dx\nonumber\\
&=& {e^{-\lambda}z\over
\beta g}dx
\label{eq: 43}
\eea
Dividing \req{eq: 43} by \req{eq: 42} and integrating for given
boundary conditions yields:
\be
ln \left({z_o\over z}\right) = {1\over \beta }\int^L_x{dx\over
g(x)}
e^{-\y (x)}
\label{eq: 44}
\ee
Using equation \req{eq: 42} to re-write the left-hand side of
\req{eq: 44} as a function of $x$ and solving for $\psi (x)$ :
\be
\psi (x) = \psi (L) - {2\over \beta} \int^L_x dx{e^{-\y (x)}\over g(x)} + ln g(L) - lng(x)
\label{eq: 45}
\ee
To find an explicit expression for $\psi (L)$ consider the
calculation of the proper time evaluated for a closed path on the boundary of the disc  at $x = L(z = z_o)$:
\be
\oint^{2\pi\beta }_{t=0}\sqrt{g(L)}dt = \oint^{2\pi }_{\theta = 0}
z_o e^{-\psi (L)/2}d\theta 
\label{eq: 46}
\ee
Integrating and solving for $\psi (L)$ :
\be
\psi (L) = -2ln\left({\beta\over z_o}\right)-lng(L)
\label{eq: 47}
\ee
Substituting \req{eq: 47} into \req{eq: 45}:
\be
\psi (x) = -lng(x) -{2\over \beta }\int^L_xdx{e^{-\y
(x)}\over g(x)}-2ln\left({\beta\over z_o}\right)
\label{eq: 48}
\ee

To solve for N we repeat the prior analysis except here we map to
a flat disc described in the form:
\be
ds^2=e^{-N(z)+{1\over \Box}\left[b(\nabla \phi)^2\right]}(z^2d\theta^2+dz^2)
\label{eq:48.01}
\ee
This results in the following:
\be
N(x)=-lng(x)-{2\over\beta}\int^L_xdx{e^{-\lambda(x)}\over g(x)}
-2ln({\beta\over z_0})+{1\over\Box}\left[b(\phi)\left(\nabla\phi (x)\right)^2
\right]
\label{eq:48.02}
\ee
Because of the non-local form of the last term this is not a 
satisfactory result so we  integrate $ \Box N = R
-b(\nabla\phi)^2 $ giving
\bea
N(x)&=& N(L)-lng(x)+lng(L)\nonumber\\
&&-\int^L_xd{\tilde x}{e^{-\lambda({\tilde x}
)}\over g({\tilde x})}\left[C-\int^L_{\tilde x}d{\overline {x}}\,b\,
e^{\lambda({\overline{x}})}(\phi^{\prime}({\overline{x}}))^2 
g({\overline{x}})\right]
\label{eq: 48.0999}
\eea
where $N(L)$ and $C$  are arbitrary constants of integration. The constant
$N(L)$ does not affect the thermodynamic quantities in the subsequent analysis,
so without loss of generality we set $N(L)=\psi(L)$. The remaining constant must in principle be determined by experiment. However, we adopt the ansatz that
$N(x)$ should reduce to $\psi(x)$ when $b=0$, and that the geometry should
uniquely determine both $\psi$ and $N$.  With these conditions
N reduces to:
\be
N(x)=\psi(x)+\int^L_xd{\tilde x}{e^{-\lambda({\tilde x})}\over
g({\tilde x})}\int^L_{\tilde x}d{\overline {x}}\,b\,e^{\lambda({\overline{x}
})}(\phi^{\prime}({\overline{x}}))^2g({\overline{x}})
\label{eq: 48.1}
\ee

If we signify $g_{CL}$ as the classical metric and $g = g_{CL} +
\delta g$ as the one-loop quantum corrected metric it can be
shown (by perturbative expansion) that the following form of the
field equation \req{eq: 35.5} is valid to first order:
\be
G_{\alpha\beta }(g) + \hbar T_{\alpha\beta }(g_{CL}) = 0
\label{eq: 49}
\ee
Where $G_{\alpha\beta }$ is given by the left hand side of
\req{eq: 8} and $T_{\alpha\beta }$ can be obtained from the
variation of \req{eq: 37}.    We find:
\bea
T_{\alpha\beta }&=&{G\over 3} \left[ \nabla_{\alpha }\nabla_{\beta
}(\psi +N) - {1\over 2}(\nabla_{\alpha}N\nabla_{\beta}\psi+
\nabla_{\alpha }\psi\nabla_{\beta }N)\right.\nonumber\\
&&\quad\quad-g_{\alpha\beta }
(R+\Box N-\half (\nabla N )\cdot(\nabla\psi)) \nonumber\\
&&\quad\quad  
-b(\psi-\ln(\mu^2))(\nabla_{\alpha}\phi\nabla_{\beta}\phi-\half g_{\alpha\beta}(\nabla\phi)^2)\nonumber\\
&&\left.\quad\quad-(g_{\alpha\beta}\Box c(\phi)-\nabla_{\alpha}
\nabla_{\beta}c(\phi))
\right]
\label{eq: 50}
\eea
An explicit expression for $ T_{\alpha\beta} $ in terms of the metric
is then obtained by substituting for $ \psi $ and N via 
\req{eq: 48} and \req{eq: 48.1} respectively. It should be noted
that the resulting equation can be equivalently obtained by direct
functional differentiation of the action in its non-local form
 (\req{eq: 36}) \cite{torre}.

We again take  the dilaton as representing the spatial
coordinate so that  the geometric corrections are manifested in the
metric.
Solving the field \req{eq: 49} yields an explicit form
of the quantum corrected metric.  In analogy to the formalism
presented by Frolov et al. \cite{frolov}  we adapt the classical static
metric (Eqs.[\ref{eq: 18}-\ref{eq: 19}]) to the quantum corrected
case as follows:
\be
ds^2 = g(x)e^{2w(x)}dt^2 + g^{-1}(x)\Omega^{-4}(\phi (x)) dx^2
\label{eq: 51}
\ee

\be
g(x) = {1\over \Omega^2(\phi (x))}\left[\jbar ({x\over l})
-2GlM - 2Glm(x)\right]
\label{eq: 52}
\ee
Here $m(x)$ is the first order quantum correction to the classical
mass $M$ and we have introduced a metric function $w(x)$ which
vanishes in the classical limit (and where functions $\jbar $ and
$\Omega^2$ are as defined by  Eqs.[\ref{eq: 12},\ref{eq:
13},\ref{eq: 16}]).

We now solve $G_{\alpha\beta } = -\hbar T_{\alpha\beta }$ by first
finding expressions for quantum quantities $m(x)$ and $w(x)$ in
terms of components of the tensor $T_{\alpha\beta }$ . Using
 \req{eq: 8} for $ G_{\alpha\beta} $ gives us:

\be
-2\nabla_{\alpha }\nabla_{\beta }D + \nabla_{\alpha
}\phi\nabla_{\beta }\phi + g_{\alpha\beta }\left[2\Box D
-\half(\nabla\phi )^2-{1\over l^2}{\tilde V}\right] = 
-\hbar T_{\alpha\beta }
\label{eq: 53}
\ee
Using the fact that the solution only depends on $x$ and the definition of
covariant derivative:
\be
-2\delta^x_{\alpha }\delta^x_{\beta
}D^{\x\x}+2\Gamma^x_{\alpha\beta }D^{\x } + \delta^x_{\alpha
}\delta^x_{\beta }(\phi^{\x})^2 + g_{\alpha\beta }
\left[2\Box D-{g^{xx}\over 2}(\phi^{\x})^2-{1\over l^2}
{\tilde V}\right] = -\hbar T_{\alpha\beta }
\label{eq: 54}
\ee
The off diagonal components (i.e. $\alpha=x$, $\beta=t$) of the above equation vanish identically. 
For the case in which both indices $\alpha ,\beta $ represent the
time coordinate: 
\be
-g^{xx}g^{\x}_{tt}D^{\x}+g_{tt}\left[2\Box D-{g^{xx}\over
2}(\phi^{\x})^2-{1\over l^2}{\tilde V}\right] = 
-\hbar T_{tt}
\label{eq: 55}
\ee
Now we re-express the left hand side with respect to the metric
defined by Eqs.[\ref{eq: 51}-\ref{eq: 52}].  First note that by using
 $D = {x\over l}$ (\req{eq: 17})  we can evaluate $\Box D$ to give

\be
\Box D = {\Omega^2\over l}(\Omega^2gw^{\x} + \jbar^{\x}-2Glm^{\x})
\label{eq: 56}
\ee
and so:
\bea
e^{2w}\left[-{2\Omega^4g^2w^{\x }\over l}-{\Omega^2g\over
l}(\jbar-2Glm)^{\x }+{g^2\Omega^2\over l}(\Omega^2)^{\x
}\right.&&\nonumber\\
\left.+{2g\Omega^2\over l}(g\Omega^2w^{\x }+\jbar^{\x }-2Glm^{\x
})- 
{\Omega^4g^2\over 2}(\phi^{\x })^2-{g\over l^2}
{\tilde V}\right] &=& -\hbar T_{tt}
\label{eq: 57}
\eea
From  Eqs.[\ref{eq: 13},\ref{eq: 16}] 
\be
\jbar^{\prime} = {1\over l}{\tilde V\over \Omega^2}
\label{eq: 58}
\ee
and from  Eqs.[\ref{eq: 12},\ref{eq: 17}] :
\be
(\Omega^2)^{\x } = {l\Omega^2\over 2}(\phi^{\x })^2
\label{eq: 59}
\ee
Using these 2 results in  \req{eq: 57} and solving for
$m^{\x }$ (and using $T_{tt} = g_{tt}T^t_t)$ :
\be
m^{\x } = {\hbar\over 2G\Omega^2}T^t_t
\label{eq: 60}
\ee
Now for the case in which both tensor indices in  \req{eq:
54} represent the spatial coordinate:
\be
-2D^{\x\x }+g^{xx}g^{\x }_{xx}D^{\x }+(\phi^{\x
})^2+g_{xx}\left[2\Box D-{g^{xx}\over 2}(\phi^{\x })^2-{1\over l^2}
{\tilde V}\right]=-\hbar T_{xx}
\label{eq: 61}
\ee
Using the metric (Eqs.[\ref{eq: 51}-\ref{eq: 52}]) along with
 \req{eq: 56} :
\bea
 &- &{1\over lg\Omega^2}(\jbar -2GlM)^{\x }-{(\Omega^2)^{\x }\over
l\Omega^2}+{(\phi^{\x })^2\over 2}\nonumber\\
&&\quad+{2\over lg\Omega^2}
(g\Omega^2w^{\x }+\jbar^{\x }-2Glm^{\x })
 - {{\tilde V}\over l^2g\Omega^4} = -\hbar T_{xx}
\label{eq: 62}
\eea
Using  Eqs.[\ref{eq: 58},\ref{eq: 59}] and solving for
$w^{\x }$ :
\be
w^{\x } = {l\over 2}\left({2G\over g\Omega^2}m^{\x }-\hbar
T_{xx}\right)
\label{eq: 63}
\ee
Substitute for $m^{\x }$ via  \req{eq: 60} and use
$T_{xx} = g_{xx}T^x_x$ :
\be
w^{\x } = {l\hbar\over 2g\Omega^4}(T^t_t - T^x_x)
\label{eq: 64}
\ee
 \req{eq: 60} and \req{eq: 64} provide the first order
quantum corrections to the geometry. Note that  consistency of
the perturbative expansion requires $T_{\alpha\beta }$ in the above 
expressions to be evaluated
on the classical solution.

  Next we explicitly evaluate the tensor
components $T^t_t$ and $T^x_x$. 
In terms of the classical static metric as expressed by 
\req{eq: 20} the non-vanishing terms are given by ( after some
simplification ):
\bea
T^t_t &=& {G\over 3} \left[\half g^{\prime}e^{2\lambda}(\psi^{\prime}
+N^{\prime})+\half g e^{2\lambda}N^{\prime}\psi^{\prime}+2 e^{\lambda}
(e^{\lambda}g^{\prime})^{\prime}\right.\nonumber\\
&&\quad\left.-b g e^{2\lambda}(\phi^{\prime})^2
(1-\half(\psi-\ln(\mu^2))-e^{2 \lambda}( gc^{\prime\prime}
+gc^{\prime}\lambda^{\prime}+\half g^{\prime}c^{\prime})
\right]
\label{eq: 65}
\eea

\bea
T^x_x&=&{G\over 3}\left[g e^{2\lambda}(\psi+N)^{\prime\prime}+(\half g^{\prime}+g\lambda^{\prime})e^{2\lambda}(\psi
+N)^{\prime}-\half g e^{2\lambda}N^{\prime}\psi^{\prime}+2 e^{\lambda}
(e^{\lambda}g^{\prime})^{\prime}\right.\nonumber\\
&&\quad\left.- b g e^{2\lambda}(\phi^{\prime})^2
(1+\half(\psi-\ln(\mu^2))-\half e^{2\lambda}g^{\prime}c^{\prime}
\right]
\label{eq: 66}
\eea
Substituting for $\psi $ (\req{eq: 48}) and $N$ (\req{eq: 48.1})  and  further simplification gives:
\bea
T^t_t  &=&{G e^{2\lambda}\over 6 g}\left[ 4g e^{-\lambda}(e^{\lambda}g^
{\prime})^{\prime}-(g^{\prime})^2+{4 \over \beta^2}e^{-2\lambda}
\right.\nonumber\\
&&\quad\quad- 2 b 
g^2(\phi^{\prime})^2\left( 1+\half ln g +{1 \over\beta}\int^L_x dx{e^
{-\lambda}\over g}+ln({\beta\over z_0\mu})\right)\nonumber\\
&&\quad\quad\left. -2{e^{-2\lambda}\over\beta}
\int^L_x dx be^{\lambda}g(\phi^{\prime})^2
 -2 g \left(gc^
{\prime\prime}+gc^{\prime}\lambda^{\prime}+\half g^{\prime}c^{\prime
}\right)\right]
\label{eq: 67}
\eea
\bea
T^x_x &=&{G e^{2\lambda}\over 6 g}\left[(g^{\prime})^2-4{e^{-2\lambda}
\over\beta^2}\right.\nonumber\\
&&\quad \quad- 
2 bg^2(\phi^{\prime})^2\left( -\half ln g-{1\over\beta}
\int^L_x dx{ e^{-\lambda}\over g}-ln({\beta\over z_0\mu})\right)\nonumber\\
&&\quad\quad\left. +2{e^{-2
\lambda}\over\beta} \int^L_x dx be^{\lambda}g(\phi^{\prime})^2
-g g^{\prime}c^{\prime}\right]  
\label{eq: 68}
\eea
Substituting these results into  \req{eq: 60} and
\req{eq: 64} gives us the desired explicit expressions for the
first order quantum corrected mass $M(x) = M_{CL}+m(x)$ and metric
function $w(x)$.  Integrating and using $\Omega^2 = e^{\y }$ leads
to the results:

\bea
M(x)&=&M_{CL}+{\hbar\over 6}\int^xdxe^{\y }\left[
2e^{-\y }(e^{\y }g^{\x })^{\x }-{(g^{\x })^2
\over 2g}+{2 e^{-2\y }\over \beta^2g}\right.\nonumber\\
&&\quad\quad- bg(\phi^{\prime})^2
\left(1+\half ln g+{1\over\beta}\int^L_x dy{e^{-\lambda(y)}\over g(y)}+ln(
{\beta\over z_0\mu})\right)\nonumber\\
&&\quad\quad\left.-{e^{-2\lambda}\over\beta g}\int^L_x dy b e^
{\lambda(y)}g(\phi^{\prime}(y))^2
-\left(gc^{\prime\prime}
+gc^{\prime}\lambda^{\prime}+\half g^{\prime}c^{\prime}\right)\right]
\label{eq: 69}
\eea

\bea
w(x) &=& -{l\hbar G\over 6}\int^L_xdx{1\over g}\left[2 e^{-\y }(e^{\y }g^{\x })^{\x }-{(g^{\x })^2\over g}+
{4 e^{-2\y }\over \beta^2g}\right.\nonumber\\
&&\quad \quad-  
2 bg(\phi^{\prime})^2\left(\half+\half ln g
+{1\over\beta}\int^L_x dy {e^{-\lambda(y)}\over g(y)}+ln({\beta\over z_0\mu})\right)
\nonumber\\ &&\quad\quad\left.
-2{e^{-2\lambda}\over\beta g}\int^L_x dy be^{\lambda(y)}g(y)(\phi^{\prime}(y))
^2
-g\left(c^{\prime\prime}+c^{\prime}\lambda^{\prime}
\right)\right]
\label{eq: 70}
\eea
Here we have imposed the condition $w(L) = 0$ and have absorbed
the lower limit of \req{eq: 69} into the constant $M_{CL}$.

Also of importance (particularly for the evaluation of the
quantum corrected entropy) is evaluation of the first order
quantum shift in the horizon and hence in the horizon value of
the dilaton field.  To this purpose we define $\Delta\phi_+ = \phi_+ - 
\phi_{+CL} $ where $\phi_+ $ and $\phi_{+CL} 
$ are the quantum corrected and classical horizon
values of the dilaton field respectively.  Because the norm of the
killing vector (\req{eq: 15}) must vanish at the horizon it follows
that the above fields must satisfy:
\be
\jbar (\phi_{+})-2M_{CL}Gl-2m(\phi_+)Gl = 0
\label{eq: 71}
\ee
\be
\jbar (\phi_{+CL})-2M_{CL}Gl = 0
\label{eq: 72}
\ee
Expanding $\jbar (\phi_+)$ about $\phi_{+CL}$ (to first order) and
using \req{eq: 58} to evaluate the derivative of $ {\overline{j}} $ gives:
\be
\jbar (\phi_+) = \jbar (\phi_{+CL})+{1\over l}\left.{\tilde
{V}(\phi_+)\over \Omega^2(\phi_+)}{1\over \phi^{\x }}
\right|_{\phi_{+CL}} \Delta\phi_+
\label{eq: 73}
\ee
Substituting for the $\jbar $'s via  Eqs.[\ref{eq:
71},\ref{eq: 72}] and solving for $\Delta\phi_+$ gives to first
order (note $m\sim\hbar $):
\be
\Delta\phi_+ = \left.{2Gl^2m(\phi )\Omega^2(\phi )\phi^{\x }\over 
\tilde {V}(\phi )}\right|_{\phi_{+CL}}
\label{eq: 74}
\ee

\section{Quantum Corrections to Black Hole Thermodynamics}

Here we calculate the thermodynamical quantities $E = (2\pi )^{-
1}\partial_{{\overline {\beta }}}W$ and $S = ({\overline {\beta }}
\partial_{{\overline {\beta }}}-1)W$ for the action functional
\req{eq: 35} which describes the one loop quantum corrected black
hole configuration:
\bea
W &=& W_{CL}\left[g\right] + \hbar\Gamma\left[g\right]\nonumber\\
  &=& W_{CL}\left[g_{CL}\right] + \hbar {\delta W_{CL}
\over \delta g}\vert_{{g_{CL}}}\delta g + \hbar\Gamma
\left[g_{CL}\right] + O\left[\hbar^2\right]
\label{eq: 75}
\eea
Recall  \req{eq: 33} for the classical action $
W_{CL}[g_{CL}]$.  This included a term with an integrand proportional
to $G^0_0$ and an inner and outer surface term.  It is possible and
convenient to derive an analogous expression for the quantum
effective action $\Gamma $.  Rewriting  \req{eq: 37} for
$\Gamma $ in terms of the static classical metric \req{eq: 20},
using  \req{eq: 25} to evaluate the extrinsic curvature
boundary term and integrating by parts leads to:
\bea
\Gamma &=&{\pi\beta\over 6}\int^L_{x_{+}}dx\left[e^{\lambda}g^{\prime}
\left(\psi +N+c\right)^{\prime}+e^{\lambda}g N^{\prime}\psi^{\prime}+b e^
{\lambda}g(\phi^{\prime})^2(\psi-\ln(\mu^2))\right]\nonumber\\
&&\quad+{\pi\over 3}\left(
\psi(x_+)+N(x_+)+c(x_+)\right)
\label{eq: 76}
\eea
Now recall  \req{eq: 65} for $ T^0_0 = T^t_t $ . Re-writing
this result ( making use of definitions of $ \Box\psi $ and $ \Box N $
and rearranging ) :
\bea
T^0_0 &=& {Ge^{\y }\over 6}\left[e^{\lambda}g^{\prime}\left(\psi+N+c\right)
^{\prime}+e^{\lambda}g N^{\prime}\psi^{\prime}+b e^{\lambda}g(\phi
^{\prime})^2(\psi-\ln(\mu^2))\right.\nonumber\\
&&\quad\left.+
4(e^{\lambda}g^{\prime})^{\prime}+2\left(e^{\lambda}g
(\psi-N)^{\prime}\right)^{\prime}-2 (e^{\lambda}gc^{\prime})^{\prime}
\right]
\label{eq: 77}
\eea
Using this result to substitute for the integrand in \req{eq: 76} 
yields :
\bea
\Gamma &=&{\pi\beta\over G}\int^L_{x_{+}}dx\left[
e^{-\lambda}T^0_0-{2 G\over 3}(e^{\lambda}g^{\prime})^{\prime}
-{G\over 3}\left(e^{\lambda}g(\psi-N)^{\prime}\right)^{\prime}
+{G\over 3}(e^{\lambda}gc^{\prime})^{\prime}\right]\nonumber\\
&&\quad\quad\quad+ 
{\pi\over 3}\left(\psi (x_+)+N(x_+)+c(x_+)\right)
\label{eq: 78}
\eea
Since three of the four terms in the integrand are total derivatives
we get :

\bea
\Gamma ={\pi\beta\over G}\int^L_{x_{+}}dx e^{-\y
}T^0_0  
&-&{2\pi\beta\over 3}e^{\y (L)}g^{\x }(L)\nonumber\\
&-&{\pi\beta\over 3}e^{\lambda(L)}g(L)\left(\psi^{\prime}(L)-
N^{\prime}(L)-c^{\prime}(L)\right)\nonumber\\
&+& 
{\pi\over 3}\left(\psi(x_+)+N(x_+)+c\phi(x_+)\right)
\label{eq: 79}
\eea
where we have used $g^{\prime
 }(x_+)e^{\y (x_+)}=2/\beta $ and 
discarded the irrelevant constant term which results.  When we
combine this result for $\Gamma $ with the first order quantum
corrected form for $W_{CL}$ into  \req{eq: 75} we obtain
an integral with integrand proportional to $G^0_0(g)+\hbar
T^0_0(g_{CL})$ along with boundary terms.  Since the integrand
vanishes on shell according to the field equation (\req{eq: 49}) we
are left with only surface contributions to $W_{on\ shell}$.  These are 
found to be

\bea
W_{on\ shell} &=& -2\pi \left[{\beta\over G}e^{\y (L)}g(L)D^{\x }(\phi
(L))+{D\over G}(\phi (x_+))+{\hbar\over 3}\beta
e^{\y (L)}g^{\x }_{CL}(L)\right.\nonumber\\
&&\quad\quad + 
{\hbar\over 6}\beta e^{\lambda(L)}g_{CL}
(L)\left(\psi^{\prime}(L)-N^{\prime}(L)-c^{\prime}(L)\right)\nonumber\\
&&\quad\quad\left.-{\hbar\over 6}\left(\psi(x_+)+N(x_+)+c_{CL}(x_+)\right)\right]
\label{eq: 80}
\eea
where the surface contributions from $W_{CL}$ are obtained by
generalizing  \req{eq: 33}.  Note that $g(x)$ and $\phi
(x_+)$ in the first two terms refer to the quantum corrected
solutions whereas the remaining terms are defined with respect
to classical geometry. Evaluation of thermodynamic quantities
is then straightforward giving:
\bea
E = - {e^{\y (L)}\over G}g^{\half }(L)D^{\x }(\phi (L))
&-&{\hbar\over 3}e^{\y (L)}g^{-\half }_{CL}(L)g^{\x }_{CL}(L)
\nonumber\\ 
&+&{\hbar\over 6}e^{\lambda(L)}g^{\half}_{CL}(L)c^{\prime}(L)
\label{eq: 81}
\eea
\be
E_{sub}=E(g;g_{CL})-E(g_0;g_{0CL})
\label{eq: 81.5}
\ee
\bea
S &=& {2\pi\over G}D(\phi (x_+))-{\hbar\over 6}2\pi
\left[2\psi(x_+)+c_{CL}(x_+)\right.\nonumber\\
\quad\quad&&+\left.\int^L_{x_{+}}dx{e^{-\lambda(x)}\over g_{CL}(x)}
\int^L_{x}d{\overline{x}}be^{\lambda({\overline{x}})}
\phi^{\prime}({\overline{x}})^2g({\overline{x}})\right]
\label{eq: 82}
\eea
Where $ g_{0} $ and $ g_{0CL} $  represent the background geometry
and are the metric fields evaluated at $ x=L\rightarrow\infty $.
Note that the left-most terms in the expressions for energy and entropy 
have classical forms but have implied quantum corrections due to geometry
. On the other hand, the remaining terms all vanish in the classical ( $ \hbar
\rightarrow 0 $ ) limit.

\section{Quantum Corrections in Spherically Symmetric Reduced
Gravity Theory}

Next we want to use the preceding formalism to examine a specific
theory.  Here we consider the form of action obtained from the
spherically symmetric
reduction of 4-dimensional Einstein-Maxwell gravity to a 2-
dimensional dilaton model \cite{SSG}. We will specifically examine the
minimal case $ b=c=0 $ so that $ N=\psi $. This case in particular
was studied by Frolov et al. \cite{frolov} and we find our results to be in agreement.

 We proceed by considering an
effective action of the following form (Note that we neglect
writing the extrinsic curvature term for sake of brevity but its
inclusion is implied.):

\be
W_{CL} = -{1\over 2Gl^2}\int d^2x\sqrt{g}\left[{r^2\over 2}R(g)+
(\nabla r)^2+\left(1-Q^2/r^2\right)\right]
\label{eq: 83}
\ee
Where $Gl^2 = G^{(4)}$ is the square of the 3+1 dimensional Planck
 length and where the ``effective'' charge $Q$ has dimensions
of length.  Comparison with the form of the classical action
(\req{eq: 13.5}) leads to the following identifications:

\be
\phi = {\sqrt{2}r\over l}
\label{eq: 84}
\ee
\be
D(\phi ) = D(r) = {r^2\over 2l^2}
\label{eq: 85}
\ee
\be
{\tilde V}(\phi ,q) = {\tilde V}(r, Q) = 1-{Q^2\over r^2}
\label{eq: 86}
\ee

The classical solution Eqs.[\ref{eq: 17}-\ref{eq: 19}] can then
be expressed:
\be
x = {r^2\over 2l}
\label{eq: 87}
\ee
\be
\Omega^2(r) = e^{\y (r)} = {r\over l}
\label{eq: 88}
\ee
\be
{\overline {j}}({x\over l}) = \jbar (r) = {r\over
l}\left(1+Q^2/r^2\right)
\label{eq: 89}
\ee
\be
g(r) = 1-{2Gl^2M\over r}+{Q^2\over r^2}
\label{eq: 90}
\ee
Note that the constant of integration in the evaluation of the 
integral defined in  \req{eq: 12} for $\Omega^2$ is
selected to be $-2\ln\sqrt{2}$. This yields 
 a metric function $g(r)$ which goes to 1 as $r\rightarrow\infty
$, which is the correct asymptotic behaviour of the metric in
spherically symmetric gravity.
For subsequent calculations it will often be convenient to express
the metric function $g(r)$ in the following form (Here we consider
solutions only for which two real, distinct horizons exist, i.e.
$(l^2GM)^2>Q^2$ .)
\be
g(r) = {1\over r^2}(r-r_+)(r-r_-)
\label{eq: 91}
\ee
where $r_{\pm }$ represents the outer(+) and inner (-) horizons
given by:
\be
r_{\pm } = l^2GM\pm\sqrt{(l^2GM)^2-Q^2}
\label{eq: 92}
\ee

Before proceeding to evaluate the quantum corrected quantities, we
consider the classical energy and entropy.  The classical energy
(\req{eq: 34}) in terms of $r$ for this theory becomes:

\bea
E &=& -{1\over G}e^{{\y }(r=L)}g^{\half }(r=L)
{dD(r)\over dr}\vert_{r=L}{dr\over dx}\nonumber\\
  &=& -{L\over Gl^2}\sqrt{1-{2Gl^2M\over L}+{Q^2\over L^2}}
\label{eq: 93}
\eea
 Above and for the remainder of this section $L$ is taken to
be the value of $r$ at the outer boundary.  Clearly this energy is
divergent as $L\rightarrow\infty $. Hence we apply the 
standard subtraction procedure as defined by \req{eq: 34.5} to give:

\be
E_{sub} = {L\over Gl^2}\left(1-\sqrt{1-{2Gl^2M\over L}+{Q^2\over
L^2}}\right)
\label{eq: 94}
\ee
Taking the asymptotic ($L \rightarrow \infty $) limit yields the
expected result $\lim_{L \rightarrow \infty }(E_{sub}) = M$.  The
classical entropy
(\req{eq: 29}) is: 
\be
S = {\pi r^2_+\over Gl^2}
\label{eq: 95}
\ee

Next we calculate the quantum corrected quantities in spherically
symmetric theory. Recall we consider the minimal case $ b=c=0 $
so that $ N=\psi $ .
  To avoid confusion classical-specific quantities
will be labelled with the subscript ``CL''.  Re-writing 
\req{eq: 69} for quantum corrected mass $M(x)$ in terms of $r$
and substituting  \req{eq: 91} for the classical metric
gives us:

\bea
M(r) &=& M_{CL}+{\hbar\over 6}\int^{r}dr\left[2{d^2g_{CL}
\over dr^2}-{({dg_{CL}\over dr})^2\over 2g_{CL}}+
{2\over \beta^2_{CL}g_{CL}}\right]\nonumber\\
     &=& M_{CL}+{\hbar\over 6}\int^rdr\left[{10\over r^4}(r-
r_{+CL})(r-r_{-CL})-{6\over r^3}(2r-r_{+CL}-r_{-CL})\right.
\nonumber\\ &&\quad+
{3\over r^2}
 - {(r-r_{-CL})\over 2r^2(r-r_{+CL})}-{(r-r_{+CL})
\over 2r^2(r-r_{-CL})}\nonumber\\
&&\quad\left.+{2r^2\over (r-r_{+CL})(r-r_{-
CL})\beta^2_{CL}}\right]
\label{eq: 96}
\eea

Integrating and using  \req{eq: 23}
\be
\beta_{CL} =\left. {2e^{-\y (r)}\over \left({dg_{CL}\over
dr}
\right)\left({dr\over dx}\right)}\right|_{r=r_{+CL}} = 
{2r^2_{+CL}\over (r_{+CL}-r_{-CL})}
\label{eq: 97}
\ee
gives us the following
\be
M(r) = M_{CL} + {\hbar\over 6}\left[A\ln(r-r_{-CL})+B\ln(r) +
C(r)\right]
\label{eq : 98}
\ee
where:
\be
A = {(r_{+CL}-r_{-CL})^2(r_{+CL} +r_{-CL})
(r^2_{+CL} +r^2_{-CL})\over 2r^4_{+CL} r^2_{-CL}}
\label{eq: 99}
\ee

\be
B = - {(r_{+CL}-r_{-CL})^2(r_{+CL}+r_{-CL})\over 2r^2_{+CL}
r^2_{-CL}}
\label{eq: 100}
\ee

\be
C(r) = {2r\over \beta^2_{CL}}+{(r_{+CL}-r_{-CL})^2\over 
2rr_{+CL}r_{-CL}}+{2(r_{+CL}+r_{-CL})\over r^2}-{10r_{+CL}
r_{-CL}\over 3r^3}
\label{eq: 101}
\ee
Note that $A+B = 4M_{CL}Gl^2/\beta^2_{CL}$.  Consider the quantum
corrected mass for
some special cases.  For $M(r = L)$ for large $L$ then
$\ln(L-r_{-CL})\sim\ln(L)$ and using
the prior property for $A$ and $B$ gives:

\be
M(L)\sim M_{CL}+{\hbar\over 3\beta^2_{CL}}[L+2M_{CL}Gl^2\ln(L)]
\label{eq: 102}
\ee
Also consider the case of an uncharged black hole. The classical
metric function becomes
$g_{CL} = (1-r_{+CL}/r)$ where $r_{+CL} = 2M_{CL}Gl^2$ and
$\beta_{CL} = 
2r_{+CL}$.  An analogous calculation to that presented above then
gives for the uncharged
case:
\be
M(r) = M_{CL}+{\hbar\over 12}\left[{r\over r_{+CL}^2}+
{7r_{+CL}\over 2r^2}-{1\over r}+{\ln(r)\over r_{+CL}}\right]
\label{eq: 103}
\ee
For the case of an extremal black hole $r_{-CL}\rightarrow
r_{+CL}$ and
$\beta_{CL}\rightarrow\infty $.  Consequently $A$ and $B$  vanish and $ C(r) $ reduces to:
\be C(r) =
4r_{+CL}/r^2-10r_{+CL}^2/3r^3
\ee

Next consider the metric function $w(x)$.  Re-writing 
\req{eq: 70} in terms of $r$
and then substituting  \req{eq: 91} for the classical
metric, using \req{eq: 97} 
for the inverse asymptotic temperature and finally integrating
gives the result

\be
w(r) = {\hbar Gl^2\over 6}\left(F(L)-F(r)\right)
\label{eq: 104}
\ee
where :

\bea
F(r) &=& -{\left[3r_{+CL}^2+2r_{+CL}r_{-CL}+3r^2_{-
CL}\right]\over r_{+CL}^2r^2_{-CL}}\ln(r)\nonumber\\
&&\quad + {\left[3r_{+CL}^4+2r_{+CL}^3r_{-CL}+2r_{+CL}^2
r^2_{-CL}+2r_{+CL}r^3_{-CL}-r^4_{-CL}\right]\over
r^4_{+CL}r^2_{-CL}}ln(r-r_{-CL})\nonumber\\
&&\quad + {4\over r^2}+{4(r_{+CL}+r_{-CL})\over rr_{+CL}r_{-CL}}
-{(r_{+CL}^4-r^4_{-CL})\over r_{+CL}^4r_{-CL}(r-r_{-CL})}
\label{eq: 105}
\eea
As for the quantum corrected mass we consider some special cases. 
For an uncharged black
hole the function $F(r)$ takes the simpler form:
\be
F(r) = {3\over 2r^2}+{2\over rr_{+CL}}-{1\over r_{+CL}^2}\ln(r)
\label{eq: 106}
\ee
If  $L$ is large, we can write:
\be
e^{2w(r)}\sim \left({r\over L}\right)^{{\hbar Gl^2\over
3r_{+CL}^2}}
\exp\left(-{\hbar Gl^2\over 3}\left({3\over 2r^2}+{2\over
rr_{+CL}}\right)\right)
\label{eq: 107}
\ee
In the extremal black hole limit the function $F(r)$ reduces to:
\be
F(r) = {8\over r_{+CL}^2}\ln
({r-r_{+CL}\over r} )+{4\over r^2}+{8\over rr_{+ CL}}
\label{eq: 108}
\ee
Consequently at the extremal black hole horizon
$F(r_+)\rightarrow -\infty $ so that 
$e^{2w(r_+)}\rightarrow\infty $. 

Next we examine the quantum corrected energy.  Revising 
\req{eq: 81} for reduced
spherically symmetric gravity:

\bea
E &=& -{L\over Gl^2}\sqrt{1-{2Gl^2M(L)\over L}+{Q^2\over
L^2}}\nonumber\\  &&\quad-{\hbar\over 3}
{1\over L^2}\left(2Gl^2M_{CL}-{2Q^2\over L} \right)
 {1\over \sqrt{1-{2Gl^2M_{CL}\over L}+{Q^2\over L^2}}}
\label{eq: 109}
\eea
Where $M(r=L)$ is given by  \req{eq: 103}.  Consider the
case of large box size
$L$.  Clearly the second part of the expression is small relative
to the first.  So we consider
the first term only and substitute for the quantum corrected mass
(for large $r = L$) by way
of  \req{eq: 102}:

\be
E\sim -{L\over Gl^2}\sqrt{1-{2Gl^2M_{CL}\over L}-{2\hbar Gl^2\over
3\beta^2_{CL}}
-{4\hbar(Gl^2)^2M_{CL}\over 3\beta^2_{CL}L}\ln(L)+{Q^2\over L^2}}
\label{eq: 110}
\ee
As in the calculation of classical energy we again apply the
standard subtraction procedure
of comparing the divergent quantity with that of a background
defined by the metric $g_0 =
\lim_{L\rightarrow\infty}g(r=L)$.  Since $g(L)$ for large $L$ is
the quantity inside the square
root sign in \req{eq: 110} it follows that $g_0 = 1 - {2\hbar
Gl^2\over 3\beta^2_{CL}}$ and
since $E[g_0] = -{Lg^{\half }_0\over Gl^2}$ the subtracted energy
is given by:

\bea
E_{sub} &\sim& {L\over Gl^2}\left[\sqrt{1-{2\hbar Gl^2\over
3\beta^2_{CL}}}\right.\nonumber\\
&&\left.- \sqrt{1-{2Gl^2M_{CL}\over L}-{2\hbar Gl^2\over 3\beta^2_{CL}}
-{4\hbar (Gl^2)^2M_{CL}\ln(L)\over 3\beta^2_{CL}L} +{Q^2\over
L^2}}\right]
\label{eq: 111}
\eea
The approach we use here is to first fix $L$ and expand the square
roots with respect to the
perturbative factor $\hbar .$  Then we take $L$ to be large and
expand with respect to
${1\over L}.$  Then eliminating all $O\left({1\over L^2}\right)$
terms inside the square
brackets leaves:

\be
E_{sub} \sim M_{CL} + {\hbar Gl^2M_{CL}\over 3\beta^2_{CL}}\left(2\ln
(L) + 1\right)
\label{eq: 112}
\ee
Note that the first order quantum correction to the energy can be
attributed to temperature effects
since  $\beta_{CL}$ represents the asymptotic inverse temperature.

Next consider the quantum corrected value of the horizon radius
$r_+$.  For this purpose we
define $\Delta r_+ = r_+-r_{+CL}$ and as previously defined $m(r)
= M(r)-M_{CL}$.  The
quantum corrected metric must vanish at $r = r_+$ and this relation
can be expressed as
follows:
\be
0 = g(r_+) = g_{CL}(r_+) - {2Gl^2m(r_+)\over r_+}
\label{eq: 113}
\ee
Expanding to first order about $r_+ = r_{+CL}$ and using
$g_{CL}(r_{+CL})=0$ and
expressing $(dg_{CL}/dr)_{r=r_{+CL}}$ in terms of $\beta_{CL}$
(\req{eq: 97}) gives:
\be
\Delta r_{+}= {\beta_{CL}G l^2 m(r_{+CL})\over r_{+CL} }
\label{eq: 114}
\ee
Note that $m(r_{+CL})$ contains a factor of $\hbar $( see \req{eq:
69}).  From this result we can
calculate the quantum correction to the  horizon area which is
proportional to $r^2_+ =
(r_{+CL} + \Delta r_+)^2$ and hence to first order:

\be
r^2_+ = r_{+CL}^2 + 2\beta_{CL}Gl^2m(r_{+CL})
\label{eq: 115}
\ee

Finally in this section we evaluate the quantum correction to the 
entropy.  For this theory the
entropy (\req{eq: 82}) is:
\be
S = {\pi r^2_+\over Gl^2} - \hbar {2\pi\over 3}\psi (r_{+CL})
\label{eq: 116}
\ee
Making use of the preceding result for $r^2_+$ (\req{eq: 115}) we can write:
\be
S = S_{CL} + 2\pi\beta_{CL}m(r_{+CL}) - \hbar {2\pi\over 3}\psi
(r_{+CL})
\label{eq: 117}
\ee
Revising \req{eq: 48} for this theory gives us:
\be
\psi (r_{+CL}) = -\ln g_{CL}(r_{+CL})-{2\over
\beta_{CL}}\int^L_{r_+}
{dr\over g_{CL}(r)}-2\ln\left({\beta_{CL}\over z_0}\right)
\label{eq: 118}
\ee
Using  \req{eq: 91} for $g_{CL}$,  \req{eq: 97}
for $\beta_{CL}$,
integrating the middle term, and simplification yields:

\bea
\psi (r_{+CL}) &=& {r^2_{-CL}\over r_{+CL}^2}\ln\left(
{L-r_{-CL}\over r_{+CL}-r_{-CL}}\right)-\ln\left(
{L-r_{+CL}\over r_{+CL}-r_{-CL}}\right)\nonumber\\
&&\quad - {(r_{+CL}-r_{-CL})\over r_{+CL}^2}(L-r_{+CL})
-2\ln\left({r_{+CL}\over z_0}\right)
\label{eq: 119}
\eea

The third term in $\psi (r_{+CL})$ can be interpreted\cite{frolov} as  the
contribution to the entropy
of a two-dimensional hot gas in a box size $L-r_{+CL}$ and
temperature 
$(2\pi\beta_{CL})^{-1} = (r_{+CL}-r_{-CL})/4\pi r_{+CL}^2$.  Hence we subtract
off this contribution to obtain the quantum corrected black hole
entropy:
\bea
S &=& S_{CL}+2\pi\beta_{CL}m(r_{+CL})-{\hbar 2\pi\over 3}{r^2_{-CL}
\over r_{+CL}^2}\ln\left({L-r_{-CL}\over
r_{+CL}-r_{-CL}}\right)\none
& &\qquad+\hbar {2\pi\over 3}\ln \left({L-r_{+CL}\over
r_{+CL}-r_{-CL}}\right)
+ \hbar {4\pi\over 3}\ln \left( {r_{+CL}\over z_0}\right)
\label{eq: 120}
\eea
We next consider some special cases.
In the case of large box size $L$ the entropy reduces to:
\be
S \sim S_{CL}+2\pi\beta_{CL}m(r_{+CL})+{2\pi\over 3}\left
(1-{r^2_{-CL}\over r_{+CL}^2}\right)\ln\left(
{L\over r_{+CL}-r_{-CL}}\right) + {4\pi\over 3}\ln
\left({r_{+CL}\over z_0}\right)
\label{eq: 121}
\ee
For an uncharged black hole then
\be
S = S_{CL} + 2\pi\beta_{CL}m(r_{+CL}) + {2\pi\over 3}
\ln \left({Lr_{+CL}\over z^2_0}\right)
\label{eq: 122}
\ee
where $m(r_{+CL})$ can be evaluated using  \req{eq:
103}.  Finally, in the
extremal black hole limit

\be
S = S_{CL} + 2\pi\beta_{CL}m(r_{+CL}) + {2\pi\over 3}\ln
\left({r_{+CL}^2\over z^2_0}\right)
\label{eq: 123}
\ee
where $m(r_{+CL}) = \hbar/9r_{+CL}$ in the extremal case however
$\beta_{CL}\rightarrow\infty $ so the entropy is divergent in this
limit.\\

\section{Quantum Corrections in Jackiw-Teitelboim Theory}

In this section we examine the
Achucarro-Ortiz black hole \cite{ortiz}, which  is a
solution to the field equations for  Jackiw-Teitelboim
gravity\cite{JT}. This theory can be obtained
by imposing axial symmetry in 2+1 dimensional gravity, so that  
the Achucarro-
Ortiz black hole corresponds to the projection of the BTZ axially
symmetric black hole \cite{BTZ}  into
1+1-dimensional spacetime. The Jackiw-Teitelboim field
equations can be derived from
an effective action of the form  \cite{ortiz}

\be
W_{CL} = -\int d^2x\sqrt{g}\Lambda^{\half }\left[rR(g) + \Lambda r
-{J^2\over 2r^3}\right]
\label{eq: 124}
\ee
where $\Lambda $ is the cosmological constant (dimension
length$^{-2}$) and $J$ is an 
``effective charge'' (dimension length) which describes the angular
momentum of the $BTZ$
black hole. Note that there is no kinetic term in this action
so it is of 
the form  (\req{eq: 13.5}) without the need for a field reparametrization. 
This leads to the following
identification (provided we set 2G = 1 and $l = \Lambda^{-\half
}$):

\be
\phibar = \Lambda^{\half }r
\label{eq: 125}
\ee

\be
D(\phibar ) = D(r) = \Lambda^{\half }r
\label{eq: 126}
\ee

\be
{\tilde V}(\phibar ,q) = {\tilde V}(r, J) = \Lambda^{\half }
(r - {J^2\over 2\Lambda r^3})
\label{eq: 127}
\ee
The classical solution Eqs.[\ref{eq: 16.25}--\ref{eq: 16.75}] can
then be expressed:

\be
x = r
\label{eq: 128}
\ee

\be
\jbar (\phibar ) = \jbar (r) = {\Lambda r^2\over 2}+{J^2\over 4r^2}
\label{eq: 129}
\ee

\be
g(r) = {\Lambda r^2\over 2}-{M\over \Lambda^{\half }}+{J^2\over
4r^2}
\label{eq: 130}
\ee
Here we consider only solutions for which two real, distinct
horizons exist (i.e. $M^2 >
\Lambda^2J^2/2$) so it will prove convenient to express the 
metric \req{eq: 130} in the
following form
\be
g(r) = {\Lambda\over 2}{(r^2-r^2_+)(r^2-r^2_-)\over r^2}
\label{eq: 131}
\ee
where $r_{+}$ and $r_-$  represent the inner and outer
horizons, respectively, given by:
\be
r^2_{\pm } = {1\over \Lambda^{3/2}}\left[M\pm 
\sqrt{M^2-\Lambda^2J^2/2}\right]
\label{eq: 132}
\ee
Note that because the action is already in reparameterized form we
set $\Omega^2 = 1$ (or
equivalently $\y = 0$) in the previously derived results.

The classical, subtracted energy for this theory,  as described by \req{eq: 34.5},  is given by:
\be
E_{sub}=2^{\half}\Lambda L\left(1-\sqrt{1-{2M\over\Lambda^{{3\over 2}
}L^2}+{J^2\over 2\Lambda L^4}}\right)
\label{eq: 133.5}
\ee
whle the  classical entropy 
(\req{eq: 29}) is:
\be
S = 4\pi\Lambda^{\half }r_+
\label{eq: 134}
\ee

Next we calculate the quantum corrected quantities in Jackiw-
Teitelboim theory with  minimal coupling ($ b=c=0 $) as for SSG in the prior
section.
Henceforth, purely classical quantities will
be labelled with the subscript ``CL.''   \req{eq: 69} for
the quantum corrected mass $M(x=r)$ gives us (substituting for
classical metric \req{eq: 131}):
\bea
M(r) &=& M_{CL}+{\Lambda\hbar\over 6}\int^rdr
\left[{4\over r^2}(r_{+CL}^2+r^2_{-CL})\right.\none
 & &\quad\left.+{5\over r^4}
(r^2-r^2_{+CL})(r^2-r^2_{-CL})
-{(2r^2-r_{+CL}^2-r^2_{-CL})\over (r^2-
r_{+CL}^2)}\right.\nonumber\\
&&\quad - \left.{(2r^2-r_{+CL}^2-r^2_{-CL})\over 
(r^2-r^2_{-CL})}+{4\Lambda^{-2}r^2\over \beta^2_{CL}(r^2-
r_{+CL}^2)(r^2-r^2_{-CL})}\right]
\label{eq: 135}
\eea
Integrating and using via  \req{eq: 23} 
\be
\beta_{CL} = {2\over \left({dg_{CL}
\over dr }\right)}\vert_{r=r_{+CL}} = {2\over \Lambda r_{+CL}
\left(1-r^2_{-CL}/r_{+CL}^2\right)}
\label{eq: 136}
\ee
gives us the following result
\be
M(r) = M_{CL}+{\hbar\Lambda\over G}\left[A\left(\ln(r-r_{-CL})-\ln
(r+r_{-CL})\right)+B(r)\right]
\label{eq: 137}
\ee
where:
\be
A = {r_{+CL}^6-3r_{+CL}^4r_{-CL}^2+3r_{+CL}^2r_{-CL}^4-r_{-CL}^6 \over
2r_{+CL}^2r_{-CL}(r_{+CL}^2-r_{-CL}^2)}
\label{eq: 138}
\ee

\be
B(r)=r+{r_{+CL}^2\over r}+{r_{-CL}^2\over r}-{5\over 3}{r_{+CL
}^2r_{-CL}^2\over r^3}
\label{eq: 139}
\ee
Next consider the quantum corrected mass for some special cases. 
For$M(r=L)$ and large $L$ then coefficient of $A\sim 0$ and we are
left with: 
\be
M(L) \sim M_{CL}+{\hbar\over 6}\Lambda L \left[ 1+{r_{+CL}^2
+r_{-CL}^2\over L^2}\right]
\label{eq: 140}
\ee
Also we consider the ``chargeless'' case.  The classical metric
function becomes $g_{CL}={\Lambda\over 2}(r^2-r_{+CL}^2)$
where $r_{+CL}^2 = 2M\Lambda^{-3/2}$ and $\beta_{CL} = 2/\Lambda
r_{+CL}$. Repeating the above calculation yields:
\be
M(r) = M_{CL} + {\hbar\over 6}\Lambda r
\label{eq: 141}
\ee
For the extremal black hole case $r_{-CL}\rightarrow r_{+CL}$ and
$\beta_{CL}\rightarrow\infty $ . The coefficient of $A$ vanishes
and $B(r)$ reduces to:
\be
B(r)=r+2{r_{+CL}^2\over r}-{5\over 3}{r_{+CL}^4\over r^3}
\label{eq: 141.5}
\ee

Next consider the metric function $w(x=r)$. Applying \req{eq:
70} for this theory and using  \req{eq: 131} for the
classical metric and  \req{eq: 136} for the inverse
asymptotic temperature gives:
\be
w(r) = {\hbar\over 6\Lambda^{\half }}(F(L)-F(r))
\label{eq: 142}
\ee
where:
\be
F(r)={4\over r}-{r(r_{+CL}^2-r_{-CL}^2)\over r_{+CL}^2(r^2-
r_{-CL}^2)}+ln({r-r_{-CL}\over r+r_{-CL}})\left[ { 3r_{+CL}^4
-2r_{+CL}^2r_{-CL}^2-r_{-CL}^4 \over 2r_{-CL}r_{+CL}^2(
r_{+CL}^2-r_{-CL}^2)}\right]
\label{eq: 143}
\ee
For the uncharged case we find using the revised metric function
discussed above that $w(r) = 0$ for all allowable $r$.  Meanwhile
for the extremal limit $F(r)$ reduces to $4/r$ except at the
horizon where the right most term in \req{eq: 143} becomes a
divergent quantity.  Hence in this limit 
$F(r_+)\rightarrow-\infty $ and as in the preceding section the
factor $e^{2w(r_+)}$ is divergent.  

Next we consider the quantum corrected energy.  Revising 
\req{eq: 81} for Jackiw-Teitelboim theory:

\be
E = -2^{\half}\Lambda L\left[\sqrt{1-{2M(L)\over\Lambda^{{3\over 2}}L^2
}+{J^2\over 2\Lambda L^4}}+{\hbar\over 3\Lambda^{\half}L}{
(1 - {J^2\over2\Lambda L^4})
\over \sqrt{1-{2M_{CL}\over \Lambda^{3\over 2 }L^2}+
{J^2\over 2\Lambda L^4}}}\right]
\label{eq: 144}
\ee
Applying the usual background subtraction procedure (\req{eq: 81.5}) then gives:
\bea
E_{sub}&=&2^{\half}\Lambda L\left[1-\sqrt{1-{2M(L)\over\Lambda^
{{3\over 2}}L^2}+{J^2\over 2\Lambda L^4}}\right.\nonumber\\
\quad\quad&&\left.+{\hbar\over 3\Lambda^{\half}L}\left(1-
{(1-{J^2\over 2\Lambda L^4})\over\sqrt{1-{2M_{CL}\over\Lambda}^{{3\over 2}}L^2}+{J^2\over 2\Lambda L^4}}\right)\right]
\label{eq: 144.5}
\eea
In the case of large box size $L$ the second part vanishes relative
to the first.  Consequently, for large $L$ the primary contribution
to the quantum shift in energy is a result of the shift in mass as
described by  \req{eq: 140}.  So to first order in $\hbar
$ and to zero'th order in ${1\over L}$ we find:
\be
E_{sub}\sim (E_{sub})_{CL} + {2^{\half }\hbar\Lambda^{\half }\over 6}
\label{eq: 145}
\ee

Following the procedure for calculating the quantum correction to the 
horizon radius $\Delta r_+ = r_+-r_{+CL}$ which was introduced in
the previous section we find
\be
\Delta r_+ = {\beta_{CL}m(r_{+CL})\over 2\Lambda^{\half }}
\label{eq: 146}
\ee
where $m(r) = M(r)-M_{CL}$ is given by  \req{eq: 137}. 
Furthermore the first order quantum correction to the  horizon area can
be obtained from:
\be
r^2_+ = r_{+CL}^2 + {\beta_{CL}r_{+CL}M(r_{+CL})\over
\Lambda^{\half }}
\label{eq: 147}
\ee

Finally, in this section we determine the quantum correction to
entropy.  For Jackiw-Teitelboim theory the dilaton generic entropy
(\req{eq: 82}) becomes:
\be
S = 4\pi\Lambda^{\half }r_+ - {\hbar 2\pi\over 3}\psi (r_{+CL})
\label{eq: 148}
\ee
Making use of the preceding result $\Delta r_+$ (\req{eq: 146}):
\be
S = S_{CL}+2\pi\beta_{CL}m(r_{+CL})-\hbar{2\pi\over 3}\psi 
(r_{+CL})
\label{eq: 149}
\ee
From  (\req{eq: 48}) we get:
\be
\psi (r_{+CL}) = -\ln g_{CL}(r_{+CL})-{2\over \beta_{CL}}
\int^L_{r_{+}}{dr\over g_{CL}(r)}-2\ln \left(
{\beta_{CL}\over z_0}\right)
\label{eq: 150}
\ee
Using  \req{eq: 131} for $g_{CL}$, \req{eq: 136} for
$\beta_{CL}$, integrating the middle term and simplifying yields:

\bea
\psi (r_{+CL}) &=& -{r_{-CL}\over r_{+CL}}\ln
\left[{(r_{+CL}-r_{-CL})(L+r_{-CL})\over (r_{+CL}+r_{-CL})
(L-r_{-CL})}\right]+\ln({r_{+CL}^2-r^2_{-CL}\over r^2_{+CL}})\none
&&\quad - \ln\left({L-r_{+CL}\over L+r_{+CL}}\right)
 +\ln\left({\Lambda z_0^2\over 8}\right)
\label{eq: 151}
\eea
So the complete quantum corrected entropy is obtained by
substituting \req{eq: 151} for $\psi (r_{+CL}$) and
$m(r_{+CL})$ via  \req{eq: 137} back into 
\req{eq: 149}.  For large $L$ this result reduces to
(subtracting off the constant term):

\bea
S &=& S_{CL}+2\pi\beta_{CL}m(r_{+CL})+{\hbar 2\pi\over 3}
{r_{-CL}\over r_{+CL}}\ln\left({r_{+CL}-r_{-CL}\over
r_{+CL}+r_{-CL}}\right)\nonumber\\
&&\quad - {\hbar 2\pi\over 3}\ln\left({r_{+CL}^2-r^2_{-CL}\over r^2_{+CL}}\right)
\label{eq: 152}
\eea
For an uncharged black hole then the entropy is given by:

\be
S = S_{CL} + \hbar {2\pi\over 3}\ln\left[{L-r_{+CL}\over 
L+r_{+CL}}\right]
\label{eq: 153}
\ee
 Note that
using \req{eq: 141} the $m(r_{+CL})$ term reduces to a constant
which we subtract off.
Finally, in the extremal black hole limit
\be
S = S_{CL}+2\pi\beta_{CL}m(r_{+CL})
\label{eq: 154}
\ee
where $m(r_{+CL}) = {2\over 9}\hbar\Lambda r_{+CL}$ in the
extremal case however $\beta_{CL}\rightarrow\infty $ so the entropy
is divergent in this limit.

\section{Conclusions}

We have calculated the one-loop quantum corrections for generic dilaton
 gravity coupled to an Abelian gauge field. Both corrections to the
 black hole geometry and black hole thermodynamics were studied in
 detail. We then applied our generic results to the special cases of
 charged black holes in spherically symmetric gravity and rotating
 BTZ black holes. The former case enabled us to verify our results
 by comparison with the tree-level calculations of Braden et al. \cite{brown}
 and the one-loop corrections of Frolov et al. \cite{frolov}. Study of BTZ
 black holes is of particular interest due to recent revelations of a 
possible 
 connection between string inspired black holes and BTZ geometry \cite{near}.

Although our quantum corrected results  can in principle be integrated
 exactly, numerical analysis will be required for rigourous study of
 particular theories. Such an analysis is in progress.
 Our hope is that ultimately such studies will lead to a better
 understanding of quantum thermodynamical
 processes associated with black holes
  and hence insight into the deep mysteries
 surrounding quantum gravity.
 
\section{Acknowledgements}
\par
 This work
was supported in part by the Natural Sciences and Engineering
Research
Council of Canada. G.K. would like to thank J. Gegenberg for helpful 
conversations. We are also grateful to S.D. Odintsov for useful comments on the original version of the manuscript and for bringing several important 
 references to our attention.
  \par

\end{document}